\documentclass[9pt,conference]{IEEEtran}
\IEEEoverridecommandlockouts
\usepackage{cite}
\usepackage{amsmath,amssymb,amsfonts}
\usepackage{algorithmic}
\usepackage[ruled]{algorithm2e}
\usepackage{graphicx}
\usepackage{textcomp}
\usepackage{xcolor}
\usepackage{hyperref}
\usepackage{colortbl}
\usepackage{multirow}
\usepackage{footnote}
\usepackage{tablefootnote}
\usepackage{float}
\usepackage{threeparttable}
 \usepackage{verbatim}
\hypersetup{
    colorlinks=true,
    linkcolor=blue,
    filecolor=magenta,      
    urlcolor=cyan,
    pdftitle={Overleaf Example},
    citecolor=blue,
}

\def\BibTeX{{\rm B\kern-.05em{\sc i\kern-.025em b}\kern-.08em
    T\kern-.1667em\lower.7ex\hbox{E}\kern-.125emX}}
    
\definecolor{Gray}{gray}{0.85}

\begin{document}
\title{Efficient yet Accurate End-to-End SC Accelerator Design}
\author{
     Meng Li$^{213*}$, Yixuan Hu$^1$, Tengyu Zhang$^1$, Renjie Wei$^1$,
    Yawen Zhang$^4$, Ru Huang$^{134}$ and Runsheng Wang$^{134*}$ 
\\
\textit{$^1$School of Integrated Circuits \& $^2$Institute for Artificial Intelligence, Peking University, China}\\
\textit{$^3$Beijing Advanced Innovation Center for Integrated Circuits, Beijing, China} \\
\textit{$^4$Institute of Electronic Design Automation, Peking University, Wuxi, China}
\vspace{-18pt}
\thanks{
This work was supported in part by the National Key R$\&$D Program of China (2020YFB2205502), NSFC (62125401) and the 111 Project (B18001). 

$^*$Corresponding author: \{meng.li, r.wang\}@pku.edu.cn}}
\maketitle
\begin{abstract}
Providing end-to-end stochastic computing (SC) neural network acceleration for state-of-the-art (SOTA) models has become an increasingly challenging task, requiring the pursuit of accuracy while maintaining efficiency.
    It also necessitates flexible support for different types and sizes of operations in models by end-to-end SC circuits.
    In this paper, we summarize our recent research on end-to-end SC neural network acceleration.
    We introduce an accurate end-to-end SC accelerator based on deterministic coding and sorting network.
    In addition, we  propose an SC-friendly model that combines low-precision data paths with high-precision residuals.
    We introduce approximate computing techniques to optimize SC nonlinear adders
    and provide some new SC designs for arithmetic operations required by SOTA models.
    Overall, our approach allows for further significant improvements in circuit efficiency, flexibility, and compatibility
    through circuit design and model co-optimization.
    The results demonstrate that the proposed end-to-end SC architecture 
    achieves accurate and efficient neural network acceleration while flexibly accommodating model requirements,
    showcasing the potential of SC in neural network acceleration.
\end{abstract}

\section{Introduction}
\label{sec:intro}
Stochastic computing (SC) has emerged as a promising alternative to traditional binary computing, 
offering simplified arithmetic operations and improved error resilience \cite{li2020acoustic, li2022sscl, zhang2020sorting, hu2022sc, ZYW2020ISCAS}.
Both hybrid and end-to-end SC-based neural accelerators have been proposed \cite{li2020acoustic, li2022sscl, zhang2020sorting, hu2022sc, ZYW2020ISCAS}.
While hybrid accelerators involve back-and-forth conversion between binary and SC representations,
leading to high power consumption and area overhead, 
end-to-end SC-based accelerators demonstrate superior power, area efficiency, and fault tolerance \cite{zhang2020sorting, ZYW2020ISCAS, hu2022sc}.
In this context, our research aims to further enhance the capabilities of end-to-end SC-based accelerators.

\begin{figure}[!tb]
    \centering
    \includegraphics[width=1\linewidth]{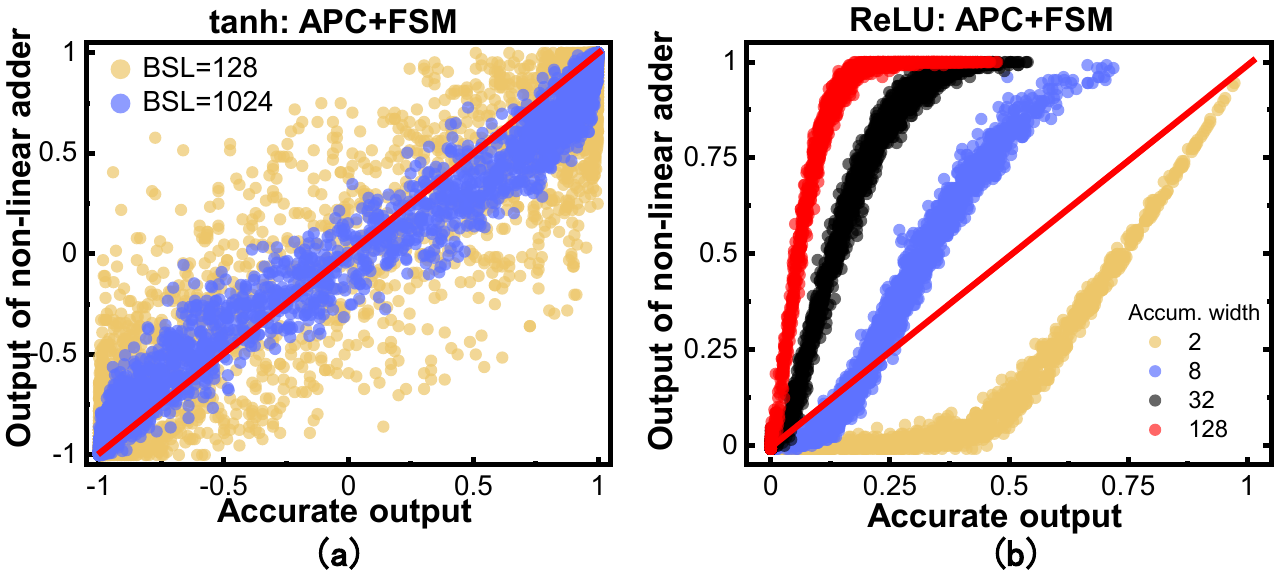}
    
    \caption{FSM-based design to implement (a) tanh and (b) ReLU. 
    Ideally, the circuit output is the same as the exact output, marked by the red line.}
    
    \label{fig:FSM}
\end{figure}

\begin{figure}[tb]
    \centering
    \includegraphics[width=0.55\linewidth]{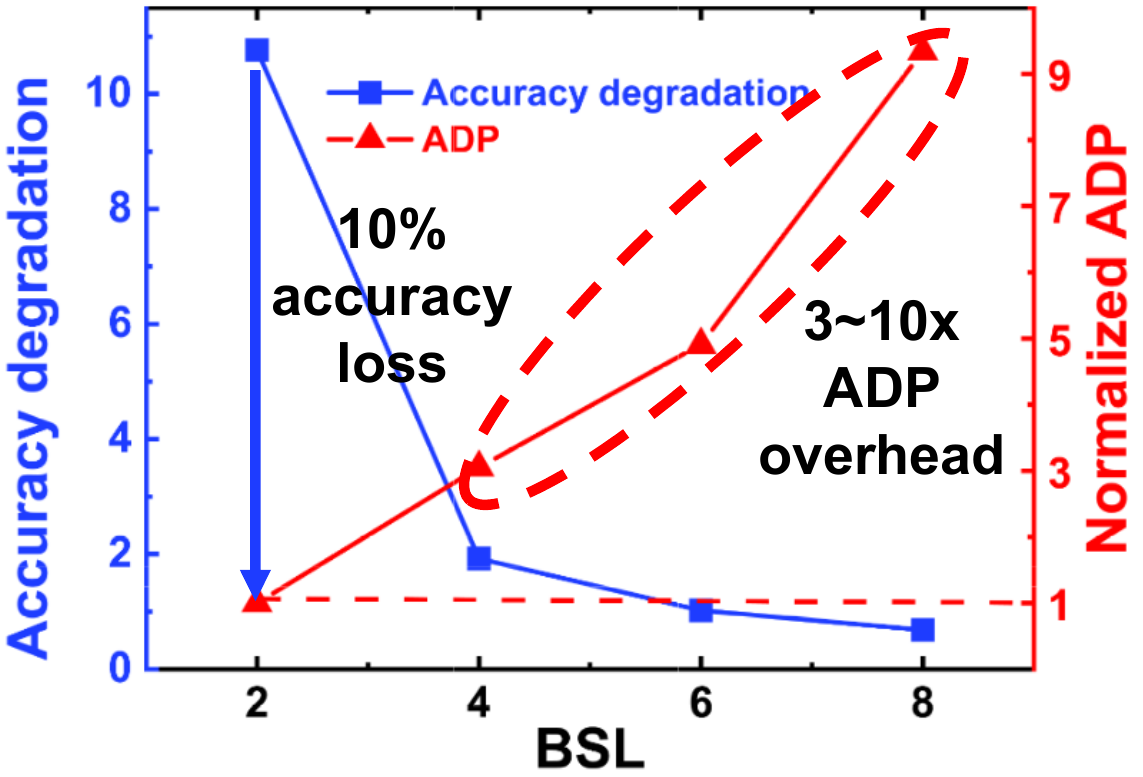}
    \caption{The trade-off between inference accuracy and efficiency (measured by area-delay product, i.e., ADP). Here, we fix the weight BSL to 2-bit and sweep the activation BSL.}
    \label{fig:tradeoff}
\end{figure}

Existing SC-based accelerators primarily focus on multiplication, accumulation, and activation functions in convolutional networks \cite{kim2016dynamic, li2017towards, li2018heif, li2020hardware}. However, these approaches have limitations. FSM-based activation modules suffer from accuracy issues, particularly for ReLU with larger accumulation widths (Figure~\ref{fig:FSM}).
Furthermore, there exists a trade-off between inference efficiency and accuracy (Figure~\ref{fig:tradeoff}), where high precision computing enhances accuracy but exponentially increases costs, while low precision computing compromises accuracy. Additionally, there is a lack of research on SC circuits supporting functions like batch normalization (BN), residual connections, gaussian error linear unit (GELU), and softmax for state-of-the-art (SOTA) models.

Therefore, in this paper, we will summarize our recent efforts on end-to-end SC-based NN accelerators that address these limitations to meet the requirements in terms of accuracy, efficiency, flexibility, and compatibility, as shown in Table~\ref{tab:accelerator_comp}. 

\begin{savenotes}
\begin{table}[!tb]
    \centering
    \caption{Comparison of different end-to-end SC accelerators.}
    \label{tab:accelerator_comp}\setlength{\tabcolsep}{1.5pt}
    \scalebox{0.88}{
\begin{threeparttable}
    \begin{tabular}{c|cccc}
    \hline \hline
    Design      & Accuracy & Efficiency & Flexibility & *Compatibility \\
    \hline
    FSM-based\cite{kim2016dynamic,li2017towards,li2018heif,li2020hardware}& Low      & Low        & Limited for large Conv & Basic CNNs \\
    Ours \cite{hu2022sc,zhang2020sorting,ZYW2020ISCAS}   & High     & Low        & Limited for variable Conv & Basic CNNs \\
    Ours \cite{HYX2023DATE, HYX2023DAC} & High     & High       & Flexible & DNNs \\
    Ours \cite{HYX2023ICCAD} & High     & High       & Flexible & DNNs+ViT \\
        \hline \hline
    \end{tabular}   
    \begin{tablenotes}
\item  \textbf{*}Basic CNNs contain convolution and ReLU. DNNs further require residual connection and BN. And transformer models further require GeLU and softmax.
    \end{tablenotes}
\end{threeparttable}}
\end{table}
\end{savenotes}
\section{Accurate End-to-end SC Accelerator Based on Deterministic Thermometer Coding}
\label{sec:sscl}

In this section, we introduce deterministic coding called thermometer coding and the corresponding end-to-end SC accelerator design. The proposed design achieves exact end-to-end SC NN acceleration.

\subsection{Motivation}
\label{subsec: motivation_sscl}

We refer to the accumulation and activation module as the SC non-linear adder. Typical SC Non-linear adders employ stochastic coding with FSM to implement different activation functions
~\cite{kim2016dynamic,li2017towards,li2018heif,li2020hardware}. 
FSM-based designs serially process stochastic bitstream inputs, which results in inaccurate outputs (Figure~\ref{fig:FSM}) that do not utilize all of the information in the inputs and have random fluctuations in the inputs themselves.
Thus, very long bitstreams, e.g., 1024 bits, are used for accuracy and lead to an unacceptable latency, which severely affects the hardware efficiency.

\subsection{Accurate End-to-End SC Acceleration with Sorting Network}
\label{subsec: method_sscl}

In our work, we employ the deterministic thermometer coding scheme (Table~\ref{tab:prec_convert}) and the corresponding accurate SC circuit designs to achieve accurate end-to-end SC acceleration.
With thermometer coding, all the 1s appear at the beginning of the bitstream and a value $x$ is represented with a $L$-bit sequence as
$x = \alpha x_q = \alpha (\sum_{i=0}^{L-1} x[i] - \frac{L}{2}),$
where $x_q = \sum_{i=0}^{L-1} x[i] - \frac{L}{2}$ is the quantized value of range $[-\frac{L}{2}, \frac{L}{2}]$ and $\alpha$ is a scaling factor obtained by training.

\begin{figure}[!tb]
    \centering
    \includegraphics[width=0.95\linewidth]{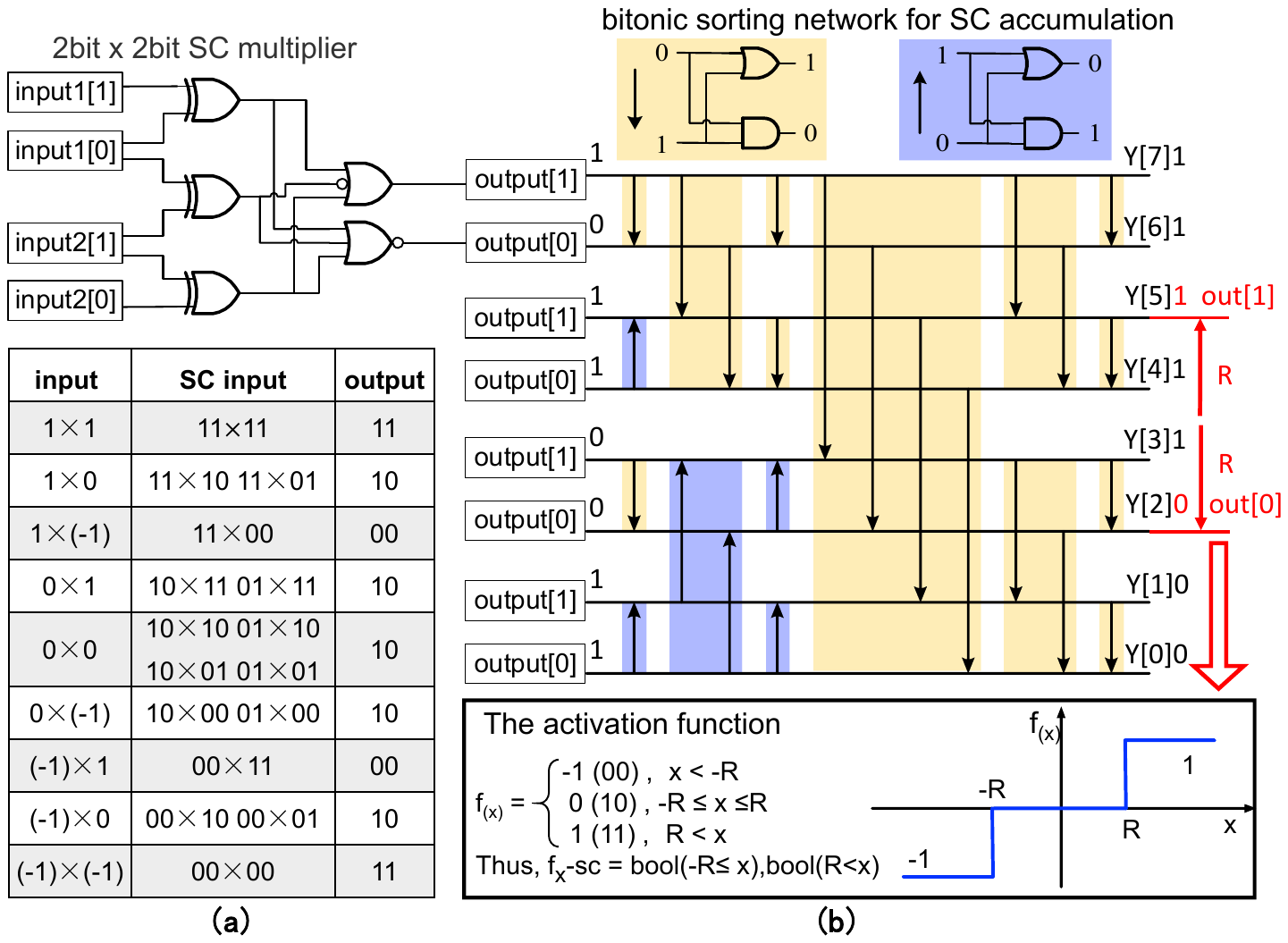}
   
    \caption{(a) The truth table and circuit of ternary SC multiplier.
    (b) The BSN and the selective interconnect system for accumulation and activation function.}
    \label{fig:bsn_intro}

\end{figure}

\begin{table}[!tb]
    \centering
    \caption{The corresponding binary precision and the represented range for thermometer coding of different BSL.}
 
    \label{tab:prec_convert}
    \resizebox{\linewidth}{!}{
    \begin{tabular}{c|c|c|c}
    \hline \hline
    BSL & \begin{tabular}[c]{@{}c@{}}Binary\\ Precision\end{tabular} & Range & Thermometer Coding \\
    \hline
    2   & - & -1, 0, 1      & 00, 10, 11  \\
    \rowcolor{Gray}
    4   & 2 & -2, -1, 0, 1, 2 & 0000, 1000, 1100, 1110, 1111 \\
    8   & 3 & -4, -3 $\cdots$ 3, 4 & 00000000, 10000000 $\cdots$ 11111110, 11111111 \\
    \rowcolor{Gray}
    16  & 4 & -8, -7 $\cdots$ 7, 8 & \begin{tabular}[c]{@{}c@{}}0000000000000000, 1000000000000000 $\cdots$\\ 1111111111111110, 1111111111111111\end{tabular} \\
    \hline \hline
    \end{tabular}
    }
\end{table}

Deterministic coding, in contrast to stochastic coding, achieves hardware-efficient and accurate computations with shorter bitstreams. By employing a 2-bit ternary bitstream, we can realize multiplication with only 5 gates using a deterministic multiplier (Figure \ref{fig:bsn_intro}(a)).

To achieve accurate accumulation and activation functions simultaneously, we employ the bitonic sorting network (BSN). BSN is a parallel sorting network that sorts inputs in thermometer coding, ensuring the output is also in thermometer coding. The sorting process, performed by comparators constructed with AND and OR gates, follows Batcher's bitonic sorting algorithm \cite{batcher1968sorting} (Figure \ref{fig:bsn_intro}(b)). The number of 1's in the sorted bitstream output from BSN corresponds to the sum of 1's in all input bitstreams, effectively representing the accumulation result.


By sorting all the bits, the inputs and outputs of the selective interconnect (SI)\cite{mohajer2018routing} are deterministic. 
Therefore, when the SI selects different bits from the BSN directly as outputs based on the selection signals, a deterministic input-output correspondence is generated and different activation functions are realized.
The example in Figure \ref{fig:bsn_intro}(b) implements the two-step activation function shown at the bottom when the SI selects the 3rd and 6th bits of the BSN as outputs.
We refer interested readers to \cite{zhang2020sorting, hu2022sc} for more details.

\subsection{Experimental Results}
\label{subsec: result_sscl}


\begin{figure}[!tb]
    \centering
    \includegraphics[width=0.9\linewidth]{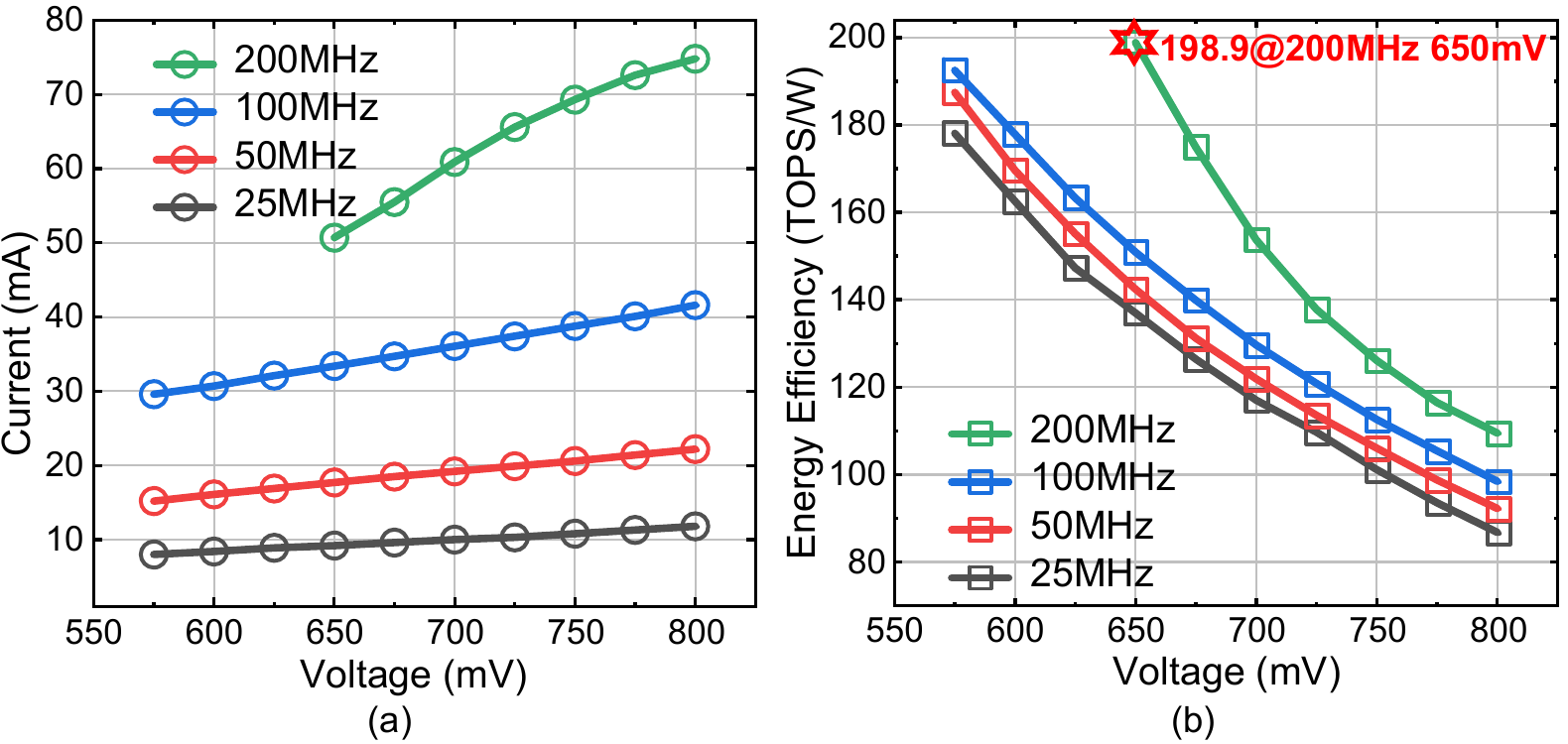}
   
    \caption{(a) Current and (b) energy efficiency versus supply voltage at different working frequencies.}
    \label{fig:chip_result}

\end{figure}

\begin{figure}[!tb]
  \centering
    \includegraphics[width=0.55\linewidth]{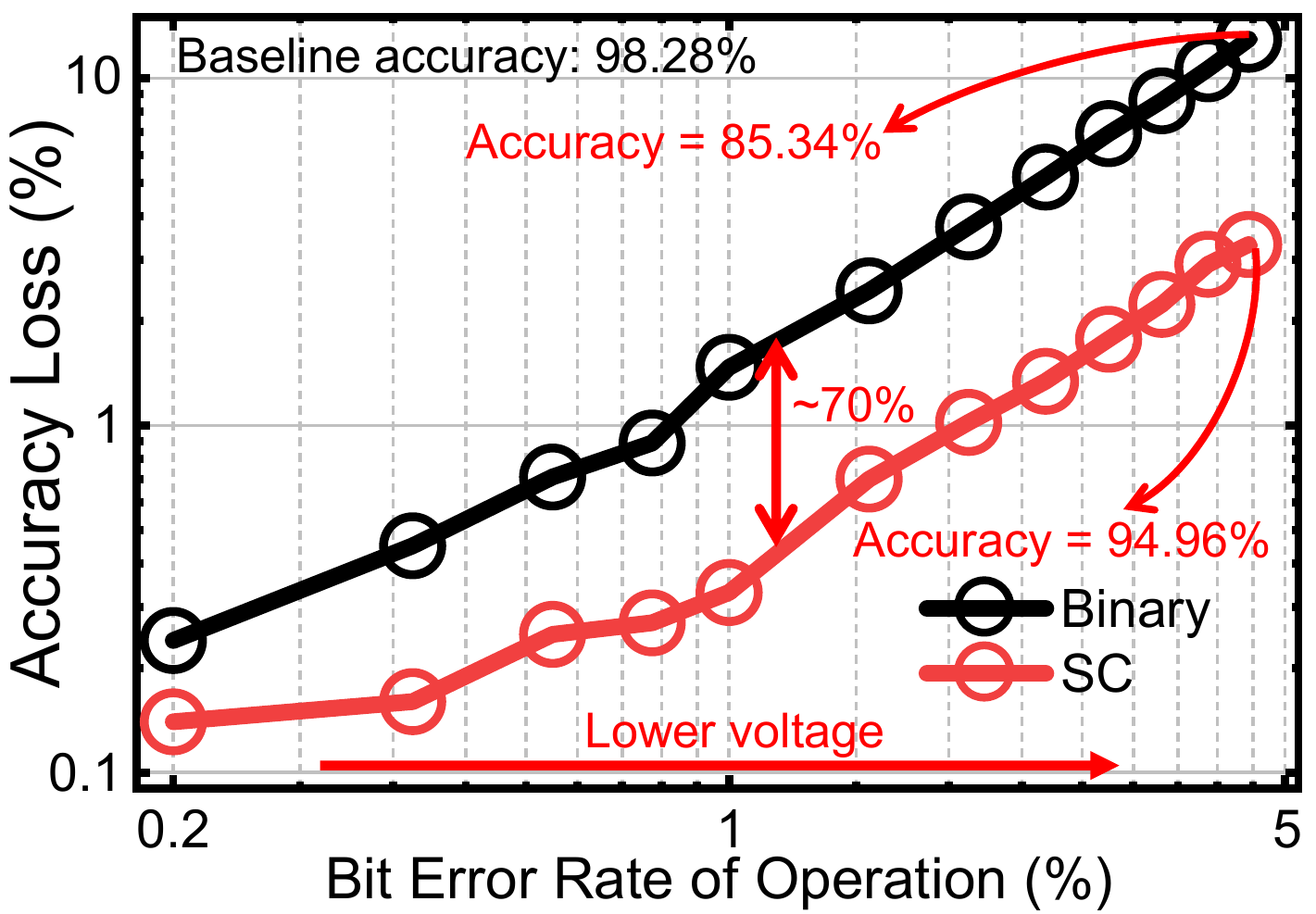}
 
    \caption{Accuracy loss of the conventional binary design and proposed SC design versus bit error rate, at the soft accuracy of 98.28\%.}
    \label{fig:accloss}
\end{figure}

We prototype the proposed SC accelerator with a 28-nm CMOS process.
The chip’s measured current consumption and energy efficiency in Figure ~\ref{fig:chip_result} show a peak of 198.9 TOPS/W at 200 MHz and 650 mV. Compared to state-of-the-art binary-based NN processors \cite{lee2018unpu, song2019, lin2020, tu2020evolver, mo2021}, the fabricated SC-based NN processor achieves an average energy efficiency improvement of 10.75$\times$ (1.16$\times\sim17.30\times$). And the area efficiency improves by 4.20$\times$ (2.09$\times\sim6.76\times$).
We also compare the accuracy under varying bit error rates (BER) using a ternary neural network that achieves 98.28\% accuracy on the MNIST dataset, 
as shown in Figure \ref{fig:accloss}. 
The proposed SC design demonstrates significant fault tolerance, as the average  reduction of accuracy loss by 70\%.
It is the first silicon-proven end-to-end SC accelerator, to the best of the authors’ knowledge.

\section{Accurate yet Efficient SC with High Precision Residual Fusion}
\label{sec:date23}

The SC accelerator above validated the effectiveness of deterministic thermometer coding and the corresponding SC design on the basic small model (MNIST). In this section, we propose SC-friendly models as well as new SC circuit blocks to support SOTA model requirements and greatly improve the accuracy of the SC accelerators.

\subsection{Motivation}
\label{subsec: motivation_date23}

The SC TNN accelerator in Section~\ref{sec:sscl} lacks support for batch normalization (BN) and residual connections, 
limiting its accuracy on complex datasets like CIFAR10 or CIFAR100. Increasing precision can enhance accuracy but compromises hardware efficiency. 
Figure \ref{fig:tradeoff} demonstrates that increasing BSL from 2 to 8 bits improves accuracy at the expense of a 3 to 10 times efficiency overhead. Accurate yet efficient SC acceleration is very challenging.

\subsection{SC-Friendly Low Precision Network}
\label{subsec: model_date23}

\begin{table}[tb]
    \centering
    \caption{Network accuracy comparison of different quantized networks on CIFAR10.}
    \label{tab:quant_acc_comp}
    \begin{tabular}{c|c c|c}
        \hline \hline
        Network              & Weight/BSL   & Act/BSL   & Top-1 Accuracy (\%) \\
        \hline
        baseline             & FP       &   FP  & 94.27  \\
        \rowcolor{Gray}
        weight quantized     & 2        &   FP  & 93.98  \\
        activation quantized &    FP    & 2     & 84.18  \\
        \rowcolor{Gray}
        fully quantized      & 2        & 2     & 83.51  \\
        \hline \hline
    \end{tabular}
\end{table}

To understand the origin of the accuracy degradation, we quantize the network weight and activation to low precision separately.
Table~\ref{tab:quant_acc_comp} shows similar accuracy between low precision weight quantization and the floating point baseline, while 2b BSL activation quantization results in a 10\% accuracy drop.
Hence, low precision activation is the root cause of the accuracy loss due to  its limited representation capacity.
After quantization, the range of activations is reduced to $\{-1, 0, +1\}$ for 2b BSL encoding, significantly reducing the number of possible configurations. 

\begin{figure}[t]
    \centering

    \includegraphics[width=0.9\linewidth]{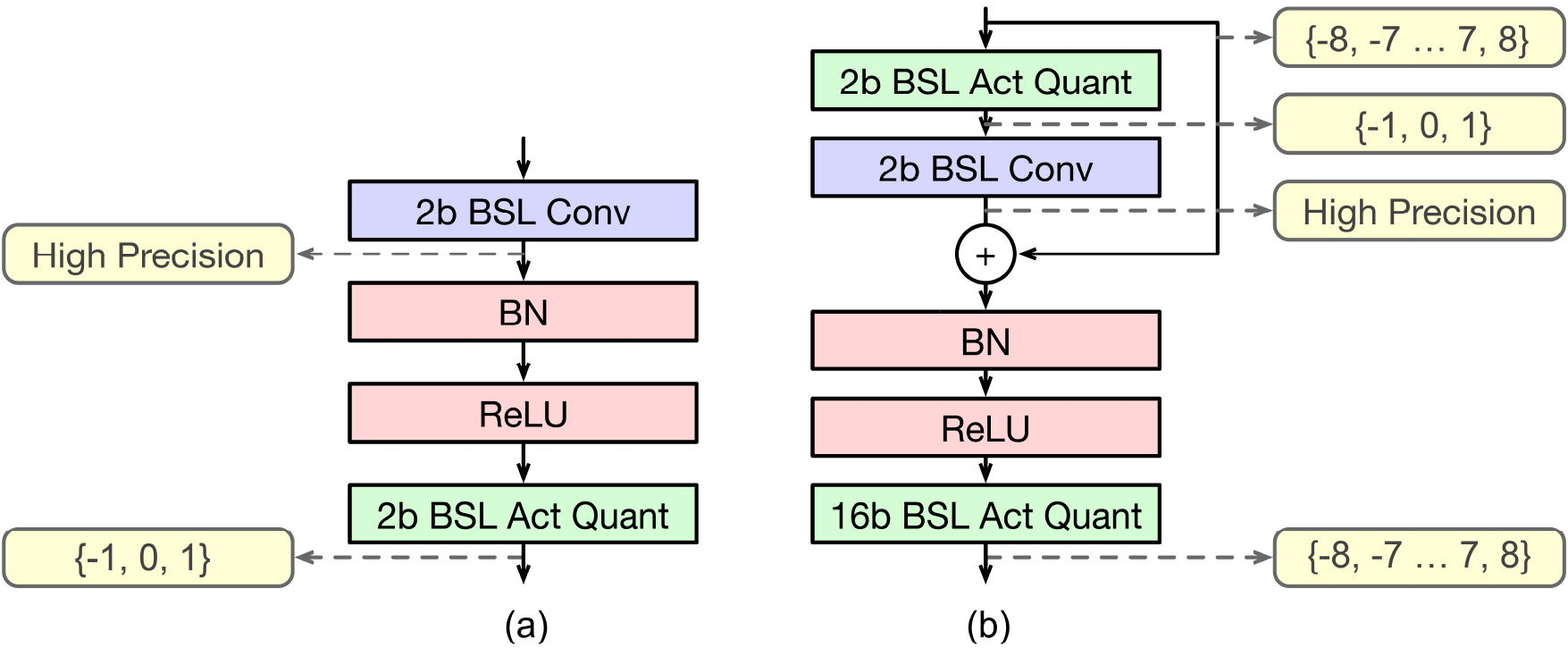}

    \caption{High precision residual helps to achieve better representation capability.}
    \label{fig:high_res}

\end{figure}

As a remedy, we add the high-precision activation input through residual connections to the result of the low-precision convolution (Figure \ref{fig:high_res}).
By increasing the activation range to $\{-8, -7, \ldots, 7, 8\}$, we enhance representation capacity to $17^{H \times W \times C}$.
This significantly improves inference accuracy 
while maintaining efficiency by preserving energy-efficient convolution computation.
\begin{align}
    \textrm{ReLU}(\textrm{BN}(x)) & = \begin{cases}
        \gamma (x - \beta) & x \geq \beta \\
        0                  & x < \beta \\
    \end{cases} \label{eq:fuse_bn}
\end{align}

Besides the high precision residual, another remaining question is how to efficiently process BN. And $BN(x) = \gamma(x - \beta)$, where $\gamma$ and $\beta$ are trainable parameters. We propose to fuse BN with the ReLU activation function as Equation~\ref{eq:fuse_bn}. Consequently, we achieve an SC-friendly low precision model with high precision residual fusion depicted in Figure~\ref{fig:high_res}(b).

\subsection{End-to-End SC Accelerator with High Precision Residual}
\label{subsec: circuit_date23}

Compared to the proposed accelerator in Section~\ref{subsec: method_sscl}, the model in Figure \ref{fig:high_res}(b) further requires the implementation of SC circuits for BN fusion and residual connection.

\begin{figure}[tb]
\centering\includegraphics[width=0.8\linewidth]{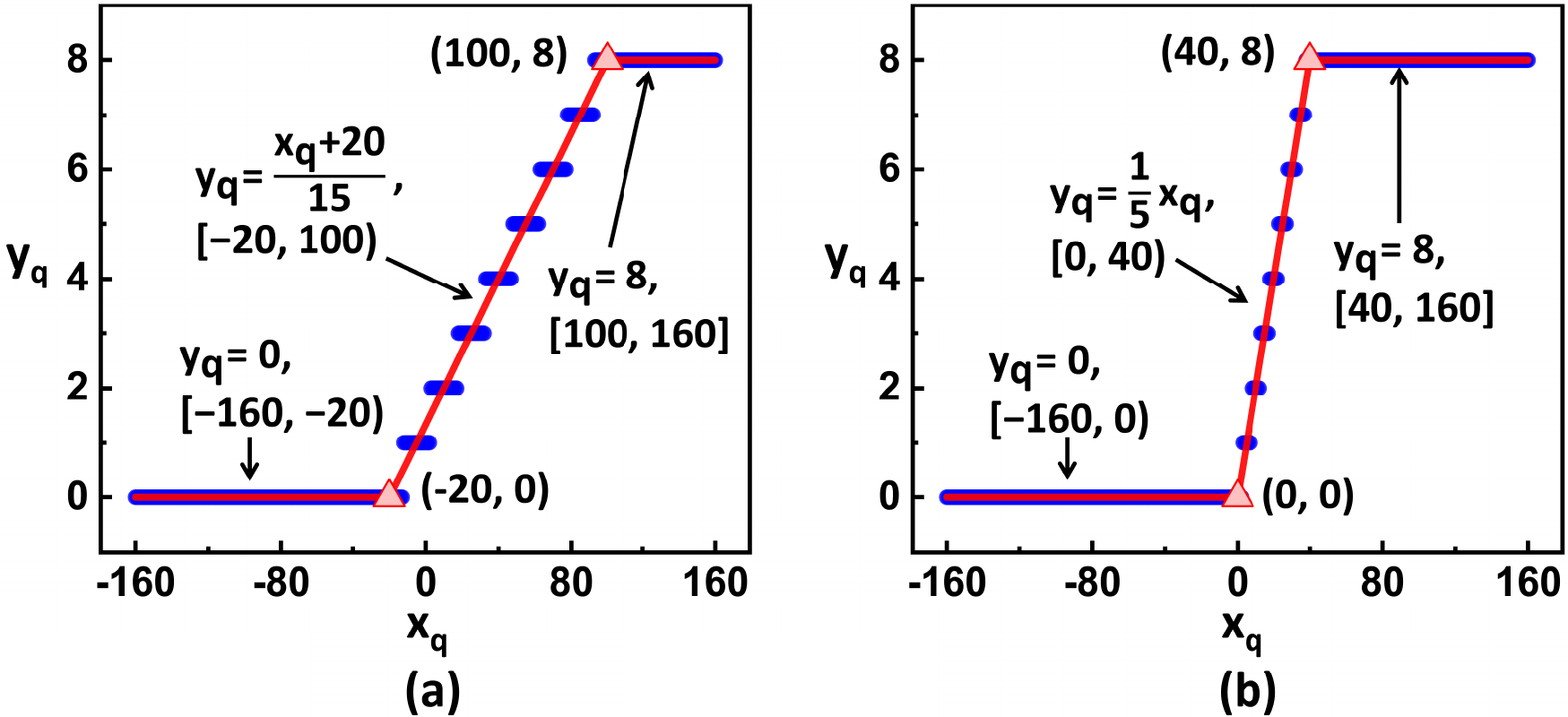}

    \caption{BN-fused activation function with 16b BSL output. The blue dots are the outputs of the proposed design for the BN-fused ReLU (Equation~\ref{eq:fuse_bn}).}
    \label{fig:bnrelu}
  
\end{figure}

The above fused BN and ReLU function can be efficiently and accurately processed in SC, leveraging the selective interconnect described in Section~\ref{subsec: method_sscl}.
Figure~\ref{fig:bnrelu} demonstrates how different BN parameters affect the objective function of the SI.

For the accumulation of residual and multiplication products, the different scaling factors $\alpha$ of residual and convolution results can lead to errors in the accumulation operation. 
The residual re-scaling block is proposed to align the $\alpha$ before accumulation. 
In the re-scaling block, we multiply or divide the residual by a factor of $2^N$ (where $N$ is an integer).
To multiply the residual by $2^N$, we replicate it $2^N$ times in the buffer. For division by $2^N$, we select 1 out of 2 bits of the residual per cycle and generate the final result after $N$ cycles. To maintain a constant BSL for the residual, we append 8 bits of '11110000' (equal to 0) per division cycle.

\subsection{Experimental Results}
\label{subsec: result_date23}

\begin{figure}[t]
\centering
  
    \includegraphics[width=0.9\linewidth]{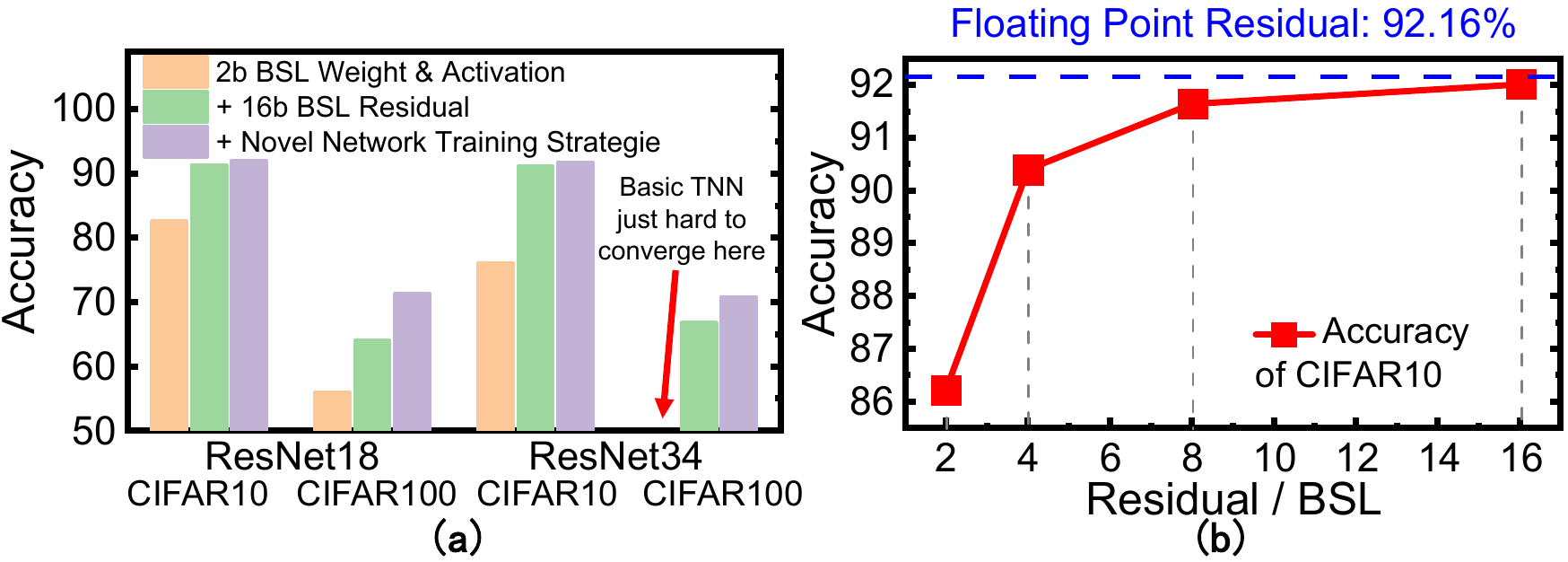}

    \caption{(a) The proposed model optimization helps to achieve much better inference accuracy; (b) 16b BSL residual achieves 5.78\% accuracy improvement, almost the same as floating point residual.}
    \label{fig:date_acc}
   
\end{figure}

Figure \ref{fig:date_acc} demonstrate significant improvement in network accuracy.  
With the high precision residual, network accuracy is improved significantly by 8.69\% and 8.12\% for low precision ResNet18 on CIFAR10 and CIFAR100, respectively. 
Combined with the novel training techniques, network accuracy can be improved in total by 9.43\% and 15.42\%.
Compared to baseline accelerators, it achieves a 9.4\% accuracy improvement with only a 1.3\% efficiency overhead compared to the efficient baseline and achieves a 3$\times$
efficiency improvement with comparable accuracy to the accurate baseline design, as shown in Table \ref{tab:acc_efficiency_comp}. In this way, the proposed method achieves accurate yet efficient SC acceleration.

\begin{table}[t]
    \centering
    \caption{Inference efficiency and accuracy comparison.}
    \label{tab:acc_efficiency_comp}
{
    \begin{tabular}{c|cc|c}
    \hline \hline
        W-A-R/BSL&
          \begin{tabular}[c]{@{}c@{}}Area (um²)\end{tabular} &
          
          \begin{tabular}[c]{@{}c@{}}ADP (um²·us)\end{tabular} &
          Accuracy (\%) \\ \hline
            2-2-2  & 4349.7   & 225.36 & 82.58 \\
        2-4-4  & 10683.3  & 687.47  & 92.35 \\
        \rowcolor{Gray}  2-2-16 & 4406.9   & 228.32  & 92.01 \\ 
        \hline \hline
        \end{tabular}
    }
\end{table}

\section{Flexible and Efficient SC Accelerator with Approximate Spatial-Temporal BSN}
\label{sec:dac23}

In this section, we greatly improve the flexibility and hardware efficiency of the SC accelerator by compressing the BSN.

\subsection{Motivation}
\label{subsec: motivation_dac23}

\begin{figure}[!tb]
    \centering
    \includegraphics[width=0.9\linewidth]{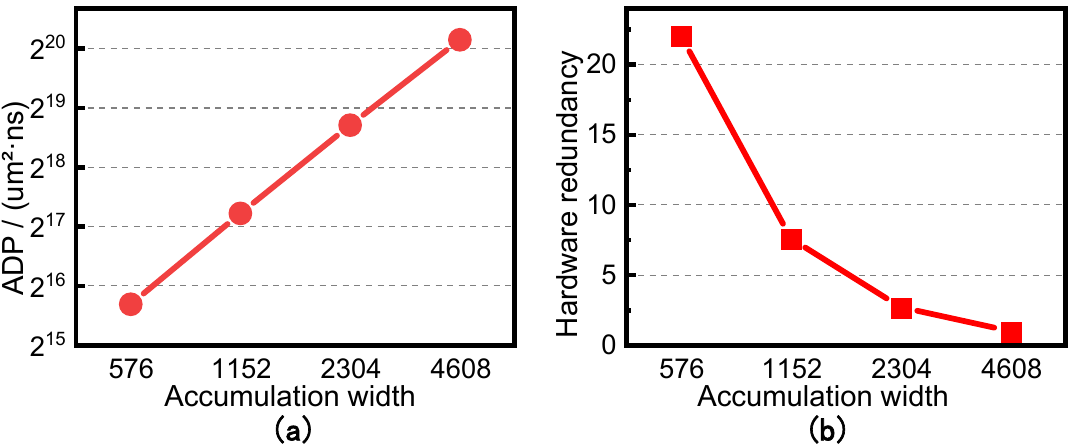}
  
    \caption{The inefficiency of the BSN design: (a) BSN hardware cost increases significantly with the accumulation widths; (b) ADP overhead using a large BSN for small accumulation widths.}
    \label{fig:bsn_adp}

\end{figure}

BSN accumulates all the input in parallel through sorting, so as to generate an accurate output based on all the information input. 
However, it also forces the hardware cost to increase super linearly with the accumulation widths (Figure~\ref{fig:bsn_adp}(a)). And the BSN has to support the largest accumulation widths among all layers. The large BSN, however, leads to very high hardware redundancy at shallow layers where the accumulation is always small (Figure~\ref{fig:bsn_adp}(b)). This makes our previous design still inefficient for SOTA models.

\subsection{Approximate Spatial-Temporal Sorting Network}
\label{subsec: method_dac23}

\begin{figure}[t]
\centering
  
\includegraphics[width=0.8\linewidth]{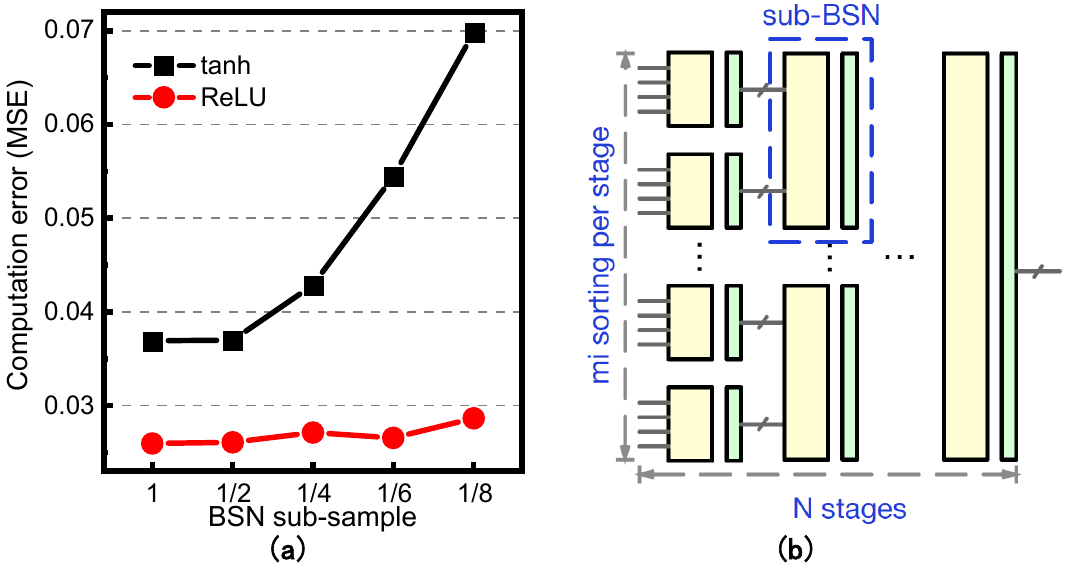}

    \caption{(a) Reducing BSN output BSL has little effect on the accuracy of SI; (b) Parameterized BSN design space.}
    \label{fig:sample}

\end{figure}

To address the inefficiency and inflexibility of BSN, we find a significant precision gap between the input and output of SI, as revealed in Figure \ref{fig:high_res}(b), making the high precision SC input redundant. We reduce the BSN output BSL, resulting in a small accuracy loss for the tanh function and negligible impact on the ReLU function, as shown in Figure \ref{fig:sample}(a).

\begin{figure}[t]
\centering\includegraphics[width=0.8\linewidth]{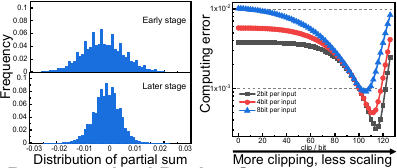}
  
    \caption{Input distribution of the intermediate sub-sampling blocks in different stages of the BSN provides an opportunity to reduce the BSN via clipping.}
    \label{fig:clip}

\end{figure}

To further reduce hardware cost, we adopt a progressive sorting and sub-sampling approach for the BSN. 
Figure~\ref{fig:sample}(b) presents a parameterized BSN design space that determines the location, number of sampling times, and method of sampling. 
The parameterized BSN consists of $N$ stages and in the $i$th stage, 
there are $m_i$ sub-BSN modules, each taking an input bitstream of $l_i$-bit BSL.
Within each sub-BSN, there is a sub-sampling block that implements truncated quantization.
It clips out $c_i$ bits on each end of the BSN while sampling 1 bit every $s_i$ bit from the remaining.
Considering the input distribution resembles a Gaussian distribution with a small variance due to inputs from a large number of multipliers, 
significant clipping can be performed with negligible errors, as illustrated in Figure~\ref{fig:clip}.

\begin{figure}[t]
\centering
 
\includegraphics[width=0.9\linewidth]{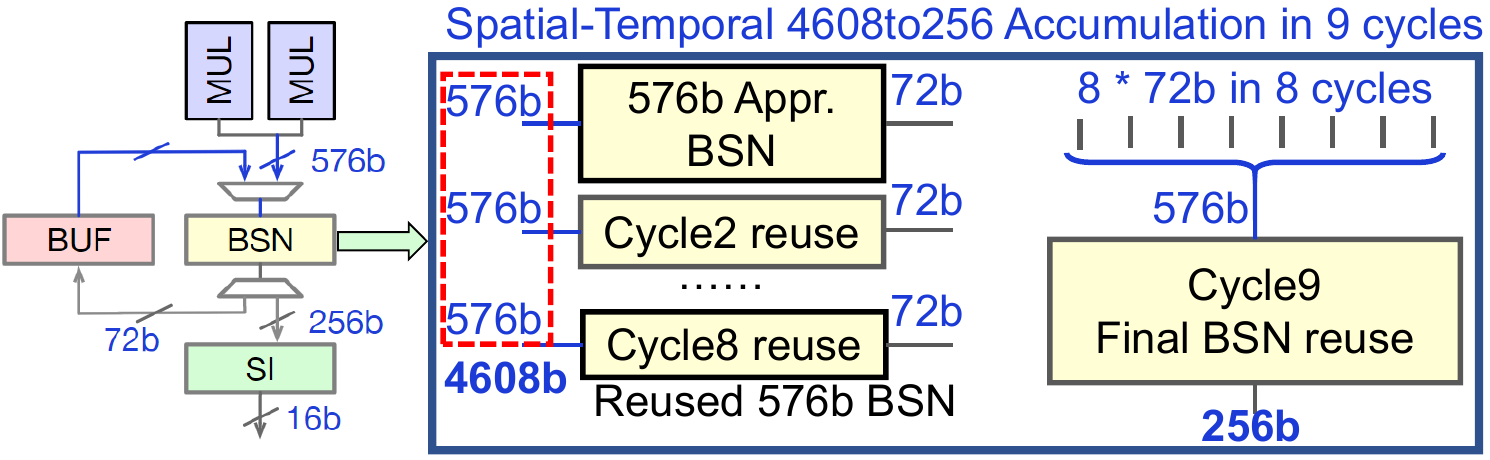}

    \caption{Spatial-temporal BSN architecture with an example: a 576-bit BSN is reused over 9 cycles for 4608b accumulation.}
    \label{fig:timefold}

\end{figure}

Thanks to the fact that the output BSL of the approximate BSN is much shorter than the input, we can further fold the accumulation temporally to achieve more flexibility. In this case, as shown in Figure~\ref{fig:timefold}, a large BSN is implemented by multi-cycle reuse of a single small BSN circuit. In the proposed spatial-temporal BSN architecture, the approximation level of BSN, i.e., the BSL of partial sums, and its corresponding reuse can be controlled through control signals. This allows for flexible handling of various accumulation widths with different approximate configurations.

\subsection{Experimental Results}
\label{subsec: result_dac23}

\begin{figure}[!tb]
    \centering
    \includegraphics[width=1\linewidth]{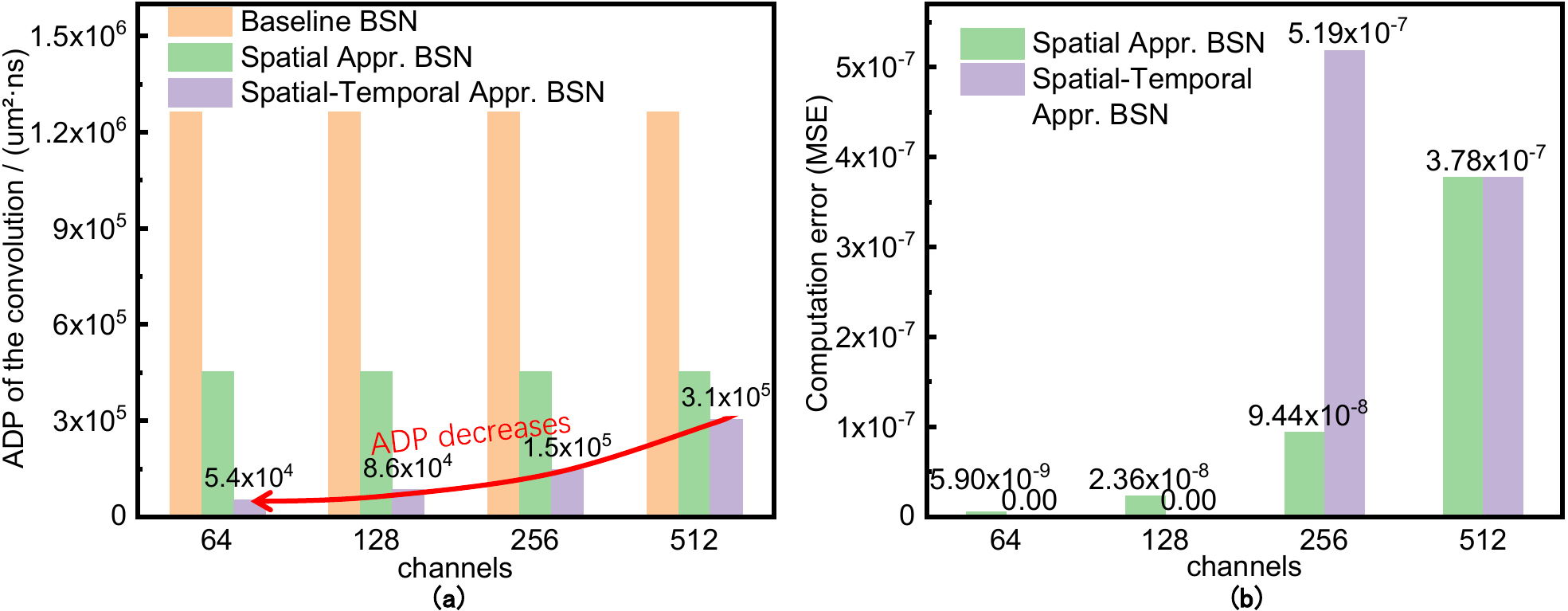}
 
    \caption{(a) ADP and (b) MSE comparison on 4 size of layers in ResNet18.}
    \label{fig:timefoldresult}
 
\end{figure}

For the largest convolution in the ResNet18, the two proposed approximate BSN reduced the ADP of BSN by 2.8$\times$ and 4.1$\times$ compared to the baseline, as shown in Table \ref{tab:time-fold}. When handling the four different sizes of convolutions in ResNet18, the spatial-temporal BSN needs fewer cycles for smaller convolutions and achieved ADP reductions from 8.2$\times$ to 23.3$\times$ with negligible errors, as shown in Figure~\ref{fig:timefoldresult}.
On average, the spatial-temporal BSN reduces the 2.2$\times$ area of datapath by reducing the average ADP of BSN by 8.5$\times$. 
This shows that the proposed SC design is more flexible and efficient.

\begin{savenotes}
\begin{table}[!tb]
    
    \centering
    \caption{Performance of different designs for a 3x3x512 convolution.}
  
    \label{tab:time-fold}
    \scalebox{0.9}{
    \begin{threeparttable}
    \begin{tabular}{c|ccc|c}
    \hline \hline
    Design & 
    Area (um²) & 
    
    Delay (ns) & 
    
    ADP (um²·ns) & 
    
    MSE     \\
    \hline
    Baseline BSN   
    & 2.95$\times10^{5}$   & 4.33  & 1.26$\times10^{6}$   & -       \\

    Spatial Appr. BSN   & 1.32$\times10^{5}$   & 3.36  & 4.55$\times10^{5}$  & 3.79$\times10^{-7}$\\
    \rowcolor{Gray} \textbf{Spatial-Temporal} & & & &
\\
\rowcolor{Gray} \textbf{Appr. BSN}  &\multirow{-2}{*}{8.18$\times10^{3}$} &\multirow{-2}{*}{1.92} & \multirow{-2}{*}{3.06$\times10^{5}$*}           & \multirow{-2}{*}{3.79$\times10^{-7}$}           \\
    \hline \hline
    \end{tabular}
    \begin{tablenotes}
       \item *Spatial-temporal BSN considers 19$\times$ area to achieve the same throughput.
    \end{tablenotes}
    \end{threeparttable}} 
\end{table}
\end{savenotes} 

\section{Summary and Future Work}
\label{sec:conclusion}
In this paper, we review our recent works on end-to-end SC neural network acceleration.
\cite{hu2022sc} implemented a parallel fully SC-based TNN processor using deterministic thermometer encoding and sorting networks on the MNIST, achieving energy efficiency of 198.9 TOPS/W. 
In addition, \cite{HYX2023DATE} propose SC-friendly models with high-precision residual fusion and corresponding SC circuits to greatly improve the network accuracy. 
\cite{HYX2023DAC} further proposed a more flexible and efficient spatial-temporal approximate BSN,
enabling accurate, efficient, and flexible end-to-end SC acceleration.
In future work, we explore SOTA transformer acceleration based on end-to-end stochastic computing, which has been submitted \cite{HYX2023ICCAD}.

\bibliographystyle{IEEEtran}
\bibliography{main}

\begin{thebibliography}{10}
\providecommand{\url}[1]{#1}
\csname url@samestyle\endcsname
\providecommand{\newblock}{\relax}
\providecommand{\bibinfo}[2]{#2}
\providecommand{\BIBentrySTDinterwordspacing}{\spaceskip=0pt\relax}
\providecommand{\BIBentryALTinterwordstretchfactor}{4}
\providecommand{\BIBentryALTinterwordspacing}{\spaceskip=\fontdimen2\font plus
\BIBentryALTinterwordstretchfactor\fontdimen3\font minus
  \fontdimen4\font\relax}
\providecommand{\BIBforeignlanguage}[2]{{%
\expandafter\ifx\csname l@#1\endcsname\relax
\typeout{** WARNING: IEEEtran.bst: No hyphenation pattern has been}%
\typeout{** loaded for the language `#1'. Using the pattern for}%
\typeout{** the default language instead.}%
\else
\language=\csname l@#1\endcsname
\fi
#2}}
\providecommand{\BIBdecl}{\relax}
\BIBdecl

\bibitem{li2020acoustic}
W.~Romaszkan \emph{et~al.}, ``ACOUSTIC: Accelerating Convolutional Neural
  Networks through Or-Unipolar Skipped Stochastic Computing,'' in \emph{Design,
  Automation \& Test in Europe Conference \& Exhibition (DATE)}, 2020, pp.
  768--773.

\bibitem{li2022sscl}
W.~Romaszkan \emph{et~al.}, ``A 4.4–75-TOPS/W 14-nm Programmable,
  Performance- and Precision-Tunable All-Digital Stochastic Computing Neural
  Network Inference Accelerator,'' \emph{IEEE Solid-State Circuits Letters},
  vol.~5, pp. 206--209, 2022.

\bibitem{zhang2020sorting}
Y.~Zhang \emph{et~al.}, ``When sorting network meets parallel bitstreams: A
  fault-tolerant parallel ternary neural network accelerator based on
  stochastic computing,'' in \emph{Design, Automation \& Test in Europe
  Conference \& Exhibition (DATE)}.\hskip 1em plus 0.5em minus 0.4em\relax
  IEEE, 2020, pp. 1287--1290.

\bibitem{hu2022sc}
Y.~Hu \emph{et~al.}, ``A 28-nm 198.9-TOPS/W Fault-Tolerant Stochastic Computing
  Neural Network Processor,'' \emph{IEEE Solid-State Circuits Letters}, vol.~5,
  pp. 198--201, 2022.

\bibitem{ZYW2020ISCAS}
Y.~Zhang \emph{et~al.}, ``Accurate and Energy-Efficient Implementation of
  Non-Linear Adder in Parallel Stochastic Computing using Sorting Network,'' in
  \emph{IEEE International Symposium on Circuits and Systems (ISCAS)}, 2020,
  pp. 1--5.

\bibitem{kim2016dynamic}
K.~Kim \emph{et~al.}, ``Dynamic energy-accuracy trade-off using stochastic
  computing in deep neural networks,'' in \emph{Proceedings of the 53rd Annual
  Design Automation Conference}, 2016, pp. 1--6.

\bibitem{li2017towards}
J.~Li \emph{et~al.}, ``Towards acceleration of deep convolutional neural
  networks using stochastic computing,'' in \emph{2017 22nd Asia and South
  Pacific Design Automation Conference (ASP-DAC)}.\hskip 1em plus 0.5em minus
  0.4em\relax IEEE, 2017, pp. 115--120.

\bibitem{li2018heif}
Z.~Li \emph{et~al.}, ``HEIF: Highly efficient stochastic computing-based
  inference framework for deep neural networks,'' \emph{IEEE Transactions on
  Computer-Aided Design of Integrated Circuits and Systems}, vol.~38, no.~8,
  pp. 1543--1556, 2018.

\bibitem{li2020hardware}
J.~Li \emph{et~al.}, ``Hardware-driven nonlinear activation for stochastic
  computing based deep convolutional neural networks,'' in \emph{2017
  International Joint Conference on Neural Networks (IJCNN)}, 2017, pp.
  1230--1236.

\bibitem{HYX2023DATE}
Y.~Hu \emph{et~al.}, ``Accurate yet Efficient Stochastic Computing Neural
  Acceleration with High Precision Residual Fusion,'' in \emph{Design,
  Automation \& Test in Europe Conference \& Exhibition (DATE)}, 2023.

\bibitem{HYX2023DAC}
Y.~Hu \emph{et~al.}, ``Efficient Non-Linear Adder for Stochastic Computing with
  Approximate Spatial-Temporal Sorting Network,'' in \emph{ACM/IEEE Design
  Automation Conference (DAC)}, 2023.

\bibitem{HYX2023ICCAD}
Y.~Hu \emph{et~al.}, ``ASCEND: Accurate yet Efficient End-to-End Stochastic
  Computing Acceleration of Vision Transformer,'' in \emph{submitted}.

\bibitem{batcher1968sorting}
K.~E. Batcher, ``Sorting networks and their applications,'' in
  \emph{Proceedings of the April 30--May 2, 1968, spring joint computer
  conference}, 1968, pp. 307--314.

\bibitem{mohajer2018routing}
S.~Mohajer \emph{et~al.}, ``Routing magic: Performing computations using
  routing networks and voting logic on unary encoded data,'' in
  \emph{Proceedings of the 2018 ACM/SIGDA International Symposium on
  Field-Programmable Gate Arrays}, 2018, pp. 77--86.

\bibitem{lee2018unpu}
J.~Lee \emph{et~al.}, ``UNPU: A 50.6 TOPS/W unified deep neural network
  accelerator with 1b-to-16b fully-variable weight bit-precision,'' in
  \emph{2018 IEEE International Solid-State Circuits Conference-(ISSCC)}.\hskip
  1em plus 0.5em minus 0.4em\relax IEEE, 2018, pp. 218--220.

\bibitem{song2019}
J.~Song \emph{et~al.}, ``7.1 An 11.5 TOPS/W 1024-MAC butterfly structure
  dual-core sparsity-aware neural processing unit in 8nm flagship mobile SoC,''
  in \emph{2019 IEEE International Solid-State Circuits
  Conference-(ISSCC)}.\hskip 1em plus 0.5em minus 0.4em\relax IEEE, 2019, pp.
  130--132.

\bibitem{lin2020}
C.-H. Lin \emph{et~al.}, ``7.1 A 3.4-to-13.3 TOPS/W 3.6 TOPS dual-core
  deep-learning accelerator for versatile AI applications in 7nm 5G smartphone
  SoC,'' in \emph{2020 ieee international solid-state circuits
  conference-(isscc)}.\hskip 1em plus 0.5em minus 0.4em\relax IEEE, 2020, pp.
  134--136.

\bibitem{tu2020evolver}
F.~Tu \emph{et~al.}, ``Evolver: A deep learning processor with on-device
  quantization--voltage--frequency tuning,'' \emph{IEEE Journal of Solid-State
  Circuits}, vol.~56, no.~2, pp. 658--673, 2020.

\bibitem{mo2021}
H.~Mo \emph{et~al.}, ``9.2 A 28nm 12.1 TOPS/W dual-mode CNN processor using
  effective-weight-based convolution and error-compensation-based prediction,''
  in \emph{2021 IEEE International Solid-State Circuits Conference (ISSCC)},
  vol.~64.\hskip 1em plus 0.5em minus 0.4em\relax IEEE, 2021, pp. 146--148.

\end{thebibliography}

\end{document}